\documentclass[prd,aps,floatfix,byrevtex,twocolumn]{revtex4}
\usepackage{epsfig}
\usepackage{graphics}
\usepackage{amsmath}
\newcommand{\be}{\begin{equation}}
\newcommand{\ee}{\end{equation}}
\newcommand{\eq}[1]{Eq.~(\ref{#1})}
\newcommand{\Dslash}{D\!\!\!\!/}
\begin{document}

 \draft
\preprint{
\vbox{
\hbox{ADP-04-11/T593}
\hbox{DESY 04-073}
}}

\title{Scaling of FLIC Fermions}

\author{J.~M.~Zanotti$^{1,2}$, B.~Lasscock$^1$, D.~B.~Leinweber$^1$ 
        and A.~G.~Williams$^1$}

\affiliation{$^1$ Special Research Centre for the
        Subatomic Structure of Matter,
        and Department of Physics,
        University of Adelaide, Adelaide SA 5005, Australia}
\affiliation{$^2$ John von Neumann-Institut f\"ur Computing
        NIC, \\
        Deutsches Elektronen-Synchrotron DESY, D-15738 Zeuthen,
        Germany}  

\begin{abstract}
  Hadron masses are calculated in quenched lattice QCD on a variety of
  lattices in order to probe the scaling behavior of the Fat-Link
  Irrelevant Clover (FLIC) fermion action, a fat-link clover fermion
  action in which the purely irrelevant operators of the fermion
  action are constructed using APE-smeared links.  The scaling
  analysis indicates FLIC fermions provide a new form of
  nonperturbative ${\cal O}(a)$ improvement where near-continuum
  results are obtained at finite lattice spacing.
\end{abstract}


\maketitle

Discretization of the continuum action of QCD for use on a space-time
lattice grid with finite lattice spacing $a$ introduces errors of
order $a^n$.  For Wilson fermions $n=1$ due to the introduction of an
irrelevant energy dimension-five lattice-Laplacian operator designed
to remove fermion doublers from the naive lattice theory by giving the
doublers a mass proportional to $a^{-1}$.  While Wilson fermions are
computationally inexpensive, the approach to the continuum limit is
slow.  Observables are spoiled by large ${\cal O}(a)$ discretization
errors and quantitative extrapolations to the continuum limit must be
performed from simulations performed at small lattice spacings
typically less than 0.1 fm.  This necessitates the use of very large
four-dimensional lattices in order to provide reasonable physical
simulation volumes exceeding 2 fm on a side.

Since the cost of simulations increases as $a^{-n}$ with $n$ exceeding
5, considerable savings have been achieved by designing lattice fermion
actions free of ${\cal O}(a)$ discretization errors.  A systematic
approach \cite{Symanzik} to achieving ${\cal O}(a)$ improvement of the
lattice fermion action in general \cite{Sheikholeslami:1985ij} is to
consider all possible gauge invariant, local dimension-five operators,
respecting the symmetries of QCD
\begin{eqnarray}
{\cal O}_1 &=& -\frac{i g a C_{SW}
  r}{4}\bar{\psi}\, \sigma_{\mu\nu}\, F_{\mu\nu}\, \psi ,\nonumber \\ 
{\cal O}_2 &=& c_2\, a\, \left\{ \bar{\psi}\, D_{\mu}\, D_{\mu}\, \psi +
  \bar{\psi}\, \overleftarrow{D}_{\mu}\, \overleftarrow{D}_{\mu}\,
  \psi\right\} , \nonumber \\ 
{\cal O}_3 &=& \frac{b_g\, a\, m_q}{2}\, {\rm tr} \left\{ F_{\mu\nu}\,
  F_{\mu\nu} \right\} 
, \label{D5operators} \\
{\cal O}_4 &=& c_4\, m_q\, \left \{\bar{\psi}\, \gamma_{\mu}\, D_{\mu}\, \psi -
  \bar{\psi}\, \overleftarrow{D}_{\mu}\, \gamma_{\mu}\, \psi \right \} ,
  \nonumber \\
{\cal O}_5 &=& -b_m\, a\, m_q^2\, \bar{\psi}\, \psi . \nonumber
\end{eqnarray}
%
%
%
Operator ${\cal O}_1$ is a new local operator in the lattice fermion
action and must be included.  On the other hand, ${\cal O}_3$ and
${\cal O}_5$ of \eq{D5operators} act to simply renormalize the
coefficients of existing terms in the lattice action, removing ${\cal
O}(a\, m_q)$ terms from the relation between bare and renormalized
quantities \cite{Dawson:1997gp}.  
%

The key observation to efficient ${\cal O}(a)$ improvement is that the
${\cal O}(a)$ improvement afforded by two-link terms of the fermion
action \cite{Heatlie:1990kg} may be incorporated to ${\cal O}(a)$ into
the standard Wilson fermion action complemented by ${\cal O}_1$ though
the following transformation of the fermion fields
\begin{eqnarray}
\psi &\rightarrow& \psi' = (1+b_q\, r\, a\, m_q)\, (1 - c_q\, r\, a\,
\Dslash )\, \psi \, , 
\nonumber \\
\bar{\psi} &\rightarrow& \bar\psi' = (1+b_q\, r\, a\, m_q)\,
\bar{\psi}\, (1 + c_q\, r\, a\, \overleftarrow{\Dslash} ) \, ,
\label{fieldrot}
\end{eqnarray}
where $\psi'$ represents the physical fermion field recovered in the
continuum limit, while $\psi$ is the lattice fermion field used in the
numerical simulations.  
At tree-level, $b_q = c_q = 1/4$ correctly incorporates the ${\cal
O}(a)$ corrections of ${\cal O}_2$ and ${\cal O}_4$ into the fermion
action.  


In summary, ${\cal O}_1$, the ``clover'' term, is the only
dimension-five operator explicitly required to complement the Wilson
action to obtain ${\cal O}(a)$ improvement.  This particular action is
known as the Sheikholeslami-Wohlert fermion
\cite{Sheikholeslami:1985ij} action
\be
S_{\rm SW} = S_{\rm W} - \frac{i g a C_{\rm SW} r}{4}\
         \bar{\psi}(x)\sigma_{\mu\nu}F_{\mu\nu}\psi(x)\ ,
\label{clover}
\ee
where $S_{\rm W}$ is the standard Wilson action \cite{Wilson}, and
$C_{\scriptstyle {\rm SW}}$ is the clover coefficient which can be
tuned to remove ${\cal O}(a)$ artifacts to all orders in the gauge
coupling constant $g$.

While this action has been known for some time, the difficulty has
been in accurately determining the renormalization of the improvement
coefficients, $C_{\rm SW}$, $b_q$, $c_q$, etc., in the interacting
quantum field theory.  At the lattice spacings typically considered in
today's lattice simulations, the tree level values can differ by a
factor of two from the renormalized values.  While mean-field improved
estimates of the renormalized coefficients provide substantial
corrections, they are not sufficiently accurate to remove ${\cal
O}(a)$ errors to all orders in the gauge coupling constant $g$
\cite{Edwards:1998nh}. 
Nonperturbative (NP) ${\cal O}(a)$ improvement \cite{Luscher:1996sc}
tunes $C_{\scriptstyle {\rm SW}}$ to all powers in $g^2$ and displays
excellent scaling \cite{Edwards:1998nh} as discussed further in the
following. 
%

In a previous paper \cite{FATJAMES}, we introduced a new form of
${\cal O}(a)$ fermion-action improvement in which the renormalization
of improvement coefficients is addressed in a very different manner.
Central to the approach is the observation that the fermion doublers
of the naive theory are removed by the Wilson term at tree level.  In
place of applying techniques to estimate the renormalization of the
improvement coefficients induced by the gauge fields of QCD,
techniques are applied to modify the gauge fields to suppress
renormalizations such that tree-level knowledge of
improvement coefficients is adequate.  

There are many accepted methods for removing short-distance
fluctuations from gauge field configurations including APE smearing,
HYP-smearing, and their variants.  The central feature of FLIC
fermions is to construct the fermion action using two sets of gauge
fields.
In the lattice operators providing the relevant dimension-four
operators of the continuum action, one works with the untouched gauge
fields generated via Monte Carlo methods, while the smoothed gauge
fields are introduced only in the purely irrelevant lattice operators
having dimension five or more.  We refer to this action as the
Fat-Link Irrelevant Clover (FLIC) fermion action.

The motivation behind constructing the FLIC action is to benefit from
the reduced exceptional configuration problem of fat-link actions
\cite{DeGrand:1998jq}, while retaining short-distance quark
interactions in the relevant operators of the fermion action.  The
expectation is that improvement in the condition number of the FLIC
fermion matrix will allow rapid calculations of fermion propagators
and efficient access to the chiral limit of full QCD
\cite{Kamleh:2004xk}.  

In this letter we present the first comprehensive scaling analysis of
FLIC fermions where four different lattice spacings are considered on
five lattices each sampled by 200 configurations.  The scaling
analysis shows convincingly that the new FLIC fermion action removes
${\cal O}(a)$ lattice artifacts from the Wilson fermion action, on a
level comparable to that of the nonperturbative improved clover
action.


The simulations are performed using an ${\cal O}(a^2)$--mean-field
improved Luscher-Weisz plaquette plus rectangle gauge action
\cite{Luscher:1984xn} on $12^3 \times 24$, $16^3 \times 32$ and $20^3
\times 40$ lattices with lattice spacings of 0.093, 0.122, 0.134 and
0.165~fm determined from a string tension analysis incorporating the
lattice coulomb potential \cite{Edwards:1998xf} with $\sqrt\sigma =
440$~MeV.
Initial studies of FLIC, mean-field improved clover and Wilson quark
actions were made using 50 configurations. The scaling analysis of
FLIC fermions presented here is performed with a total of 200
configurations at each lattice spacing and volume.
Gauge configurations are generated using the Cabibbo-Marinari
pseudo heat-bath algorithm with three diagonal SU(2) subgroups looped
over twice.  Simulations are performed using a parallel algorithm with
appropriate link partitioning \cite{Bonnet:2000db}, and the error
analysis is performed by a third-order, single-elimination jackknife,
with the $\chi^2$ per degree of freedom ($N_{\rm DF}$) obtained via
covariance matrix fits.


Fat links \cite{DeGrand:1998jq,DeGrand:1999gp} are created by
averaging or smearing links on the lattice with their nearest
transverse neighbors in a gauge covariant manner (APE smearing).  The
smearing procedure \cite{APE} replaces a link, $U_{\mu}(x)$, with a
sum of the link and $\alpha$ times its staples
\begin{eqnarray}
&&\!\!U_{\mu}(x) \rightarrow U_{\mu}'(x) =
(1-\alpha) \, U_{\mu}(x) \\
&&\!\!+ \frac{\alpha}{6}\sum_{\nu=1 \atop \nu\neq\mu}^{4}
  \Big[ U_{\nu}(x)\,
        U_{\mu}(x+\nu a)\,
        U_{\nu}^{\dag}(x+\mu a)                         \nonumber \\
&&\!\!+ U_{\nu}^{\dag}(x-\nu a)\,
        U_{\mu}(x-\nu a)\,
        U_{\nu}(x-\nu a +\mu a)
  \Big] \,,  \nonumber
\end{eqnarray}
followed by projection back to SU(3). We select the unitary matrix
$U_{\mu}^{\rm FL}$ which maximizes
\be
{\cal R}e \, {\rm{tr}}(U_{\mu}^{\rm FL}\, U_{\mu}'^{\dagger})\, ,
\ee
by iterating over the three diagonal SU(2) subgroups of SU(3).
Performing eight iterations over these subgroups gives gauge
invariance up to seven significant figures.  The combined procedure of
smearing and projection is repeated to create a fat link.

The mean-field improved FLIC action now becomes
\be
S_{\rm SW}^{\rm FL}
= S_{\rm W}^{\rm FL} - \frac{i\, g\, C_{\rm SW}\, \kappa\, r}{2(u_{0}^{\rm FL})^4}\
             \bar{\psi}(x)\sigma_{\mu\nu}F_{\mu\nu}\psi(x)\ ,
\label{FLIC}
\ee
where $F_{\mu\nu}$ is constructed using fat links, $u_{0}^{\rm FL}$ is
the mean link calculated with fat links,
and where the mean-field improved Fat-Link Irrelevant Wilson action is
\begin{eqnarray}
S_{\rm W}^{\rm FL}
=  &\sum_x& \bar{\psi}(x)\psi(x) 
+ \kappa \sum_{x,\mu} \bar{\psi}(x)
    \bigg[ \gamma_{\mu}
      \bigg( \frac{U_{\mu}(x)}{u_0} \psi(x+\hat{\mu}) \nonumber \\
&-& \frac{U^{\dagger}_{\mu}(x-\hat{\mu})}{u_0} \psi(x-\hat{\mu})
      \bigg)
- r \bigg(
 \frac{U_{\mu}^{\rm FL}(x)}{u_0^{\rm  FL}} \psi(x+\hat{\mu})\nonumber\\
&+& \frac{U^{{\rm FL}\dagger}_{\mu}(x-\hat{\mu})}{u_0^{\rm FL}}
          \psi(x-\hat{\mu})
      \bigg)
    \bigg]\ .
\label{MFIW}
\end{eqnarray}
with $\kappa = 1/(2m+8r)$. We take the standard value $r=1$.  Our
notation uses the Pauli (Sakurai) representation of the Dirac
$\gamma$-matrices defined in Appendix~B of Sakurai \cite{Sakurai}. In
particular, the $\gamma$-matrices are hermitian and $\sigma_{\mu\nu} =
[\gamma_{\mu},\ \gamma_{\nu}]/(2i)$.

As reported in Ref.~\cite{FATJAMES}, the mean-field improvement
parameter for the fat links is very close to 1.  Hence, the mean-field
improved coefficient for $C_{\rm SW}$ is expected to be accurate.
A significant advantage of the fat-link irrelevant operator approach
is that one can now use highly improved definitions of $F_{\mu\nu}$
which give impressive near-integer results for the topological charge
\cite{sbilson}.

In particular, we employ the 3-loop ${\cal O}(a^4)$-improved definition of
$F_{\mu\nu}$ in which the standard clover-sum of four $1 \times 1$
loops lying in the $\mu ,\nu$ plane is combined with $2 \times 2$ and $3
\times 3$ loop clovers.
Bilson-Thompson {\it et al.} \cite{sbilson} find
\begin{eqnarray}
g\, F_{\mu\nu} &=& \frac{1}{8i} \left [ \left ( 
 \frac{3}{2}      W^{1 \times 1} -
 \frac{3}{20\, u_0^4}W^{2 \times 2} +
 \frac{1}{90\, u_0^8}W^{3 \times 3}\right ) \right .
\nonumber  \\
 &&\qquad -\ {\rm h.c.} \Big ]_{\rm traceless}
\label{Fmunu}
\end{eqnarray}
where $W^{n \times n}$ is the clover-sum of four $n \times n$ loops
and $F_{\mu\nu}$ is made traceless by subtracting $1/3$ of the trace from
each diagonal element of the $3 \times 3$ color matrix.
This definition reproduces the continuum limit with ${\cal O}(a^6)$
errors.
On approximately self-dual configurations, this operator produces
integer topological charge to better than 4 parts in $10^4$.  We have
also considered a 5-loop improved $F_{\mu\nu}$
which agrees with the 3-loop version to better than 4 parts in $10^4$
\cite{sbilson}.

An important consideration is the amount of smearing to apply to the
gauge fields of the irrelevant operators.  
%
%
Since the aim is to remove perturbative renormalizations of the
improvement coefficients, we monitor the 3-loop ${\cal
O}(a^4)$-improved topological charge constructed with the 3-loop
${\cal O}(a^4)$-improved definition of $F_{\mu\nu}$ of
Eq.~(\ref{Fmunu}) as a function of smearing sweep.  The topological
charge is known to have a large multiplicative renormalization
\cite{campostrini} and serves as an ideal operator for monitoring the
removal of perturbative physics under smearing.  We find that the
topological charge varies rapidly over the first few sweeps of
smearing but then makes only small variations thereafter.
We define the optimal number of sweeps to be the minimum number of
sweeps required to reach the smoothly varying regime.
It is interesting that our findings for optimal smearing coincide with
that required to provide the optimal condition number of the FLIC
fermion matrix in the negative mass regime relevant to overlap
fermions \cite{Kamleh:2001ff}.


\begin{figure}[t]
\begin{center}
{\includegraphics[angle=90,width=0.98\hsize]{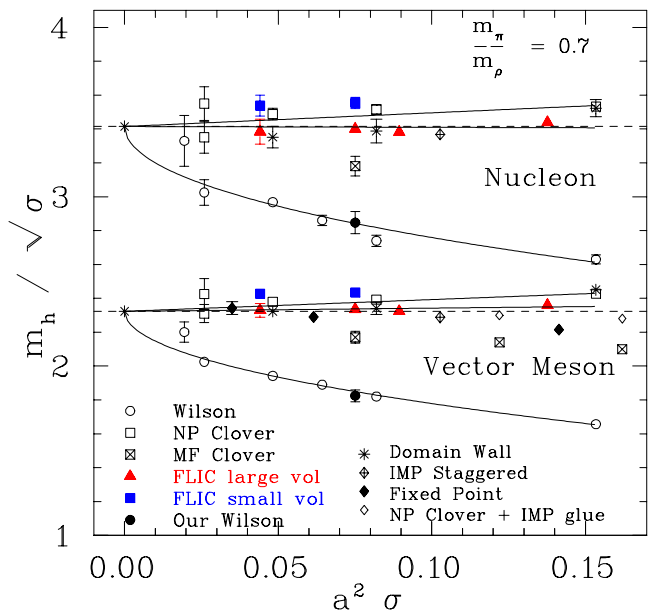}}
\caption{Nucleon and vector meson masses for the Wilson, Mean-Field
  (MF) improved, NP-improved clover, domain wall, fixed point,
  improved staggered and FLIC actions obtained by interpolating
  simulation results to $m_\pi / m_\rho = 0.7$.  Results from the
  current simulations are indicated by the solid symbols; those from
  earlier simulations by open or hatched symbols.  The solid-lines
  illustrate fits, constrained to have a common continuum limit, to
  FLIC, NP-improved clover and Wilson fermion action results obtained
  on physically large lattice volumes.  Further details are provided
  in the text.}
\label{scaling1} 
\end{center}
\vspace{-0.5cm}
\end{figure}

Hadron masses are extracted from the Euclidean time dependence of the
calculated two-point correlation functions using standard techniques
\cite{Spin32}. 
To compare with the results of Ref.~\cite{Edwards:1998nh}, we
interpolate our results to a pseudo-scalar to vector meson mass ratio
of $m_\pi / m_\rho = 0.7$.  The scaling behavior of the various
fermion actions is illustrated in Fig.~\ref{scaling1}.  Lattice
volumes, spacings and hadron masses in units of the string tension are
given in Table~\ref{lattices}.

Since the smearing radius \cite{Bonnet:2001rc} is proportional to the
product of $\alpha$ and the number of smearing sweeps, $n$, we fix
$\alpha = 0.7$ and vary $n$.  For the fine lattices with $a^2 \sigma
\sim 0.075$ and 0.045, four smearing sweeps are performed.  For the
coarser lattices with $a^2 \sigma \sim 0.09$ and 0.14 we perform six
and eight smearing sweeps respectively.  
The effective range of the smearing \cite{Bernard:1999kc} within
which interactions are suppressed is $\langle r^2 \rangle = a^2\,
\alpha\, n / 3$, providing an RMS radius of 1.0, 1.2 and 1.4 lattice
spacings for $n = 4$, 6 and 8 smearing sweeps at $\alpha = 0.7$
respectively.

Actions with fat-link irrelevant operators perform extremely well,
lying very near the horizontal dashed lines corresponding to continuum
limit results at finite lattice spacing.
For reference we have also calculated masses for the Wilson action at
$a^2 \sigma \simeq 0.075$ which agree with those of
Ref.~\cite{Edwards:1998nh}.
We also compare our results with the standard Mean-Field Improved
Clover (MFIC) action.  We mean-field improve as defined in
Eqs.~(\ref{FLIC}) and (\ref{MFIW}) but with thin links throughout. For
this action, the standard 1-loop definition of $F_{\mu\nu}$ is used.
For both the vector meson and the nucleon, the FLIC actions perform
significantly better than the mean-field improved clover action.

Finally, our FLIC results compare extremely well with a variety of
improved actions found in the literature. In particular, in
Fig.~\ref{scaling1} we compare with results using domain wall
\cite{DW}, fixed point \cite{FP} and improved staggered fermions
\cite{MILC}.

\begin{table}[t]
\begin{center}
\caption{Lattice parameters
  and results for the vector
  meson and nucleon masses interpolated to $m_\pi / m_\rho = 0.7$. 
  \label{lattices}}
\begin{ruledtabular}
  \begin{tabular}{ccccccc}
        $\beta$ & Volume & $N_{\rm configs}$ & $a\sqrt{\sigma}$ & 
        $m_v / \sqrt{\sigma}$ & $m_N / \sqrt{\sigma}$ & $u_0$     \\ \hline
    4.38  & $16^3\times 32$ & 200 & 0.371 & 2.360(20) & 3.439(27) & 0.8761 \\
    4.53  & $20^3\times 40$ & 200 & 0.299 & 2.324(15) & 3.382(24) & 0.8859 \\
    4.60  & $12^3\times 24$ & 200 & 0.274 & 2.434(26) & 3.554(33) & 0.8889 \\
    4.60  & $16^3\times 32$ & 200 & 0.274 & 2.336(22) & 3.400(26) & 0.8889 \\
    4.80  & $16^3\times 32$ & 200 & 0.210 & 2.427(23) & 3.538(61) & 0.8966 \\
\end{tabular}
\end{ruledtabular}
\end{center}
\vspace{-0.5cm}
\end{table}

The two different volumes used at $a^2 \sigma \sim 0.075$ reveal a
small finite volume effect, which increases the mass for the smaller
volumes at $a^2 \sigma \sim 0.075$ and $\sim 0.045$.  
Examination of points from the small and large volumes separately
indicates continued scaling toward the continuum limit. While the
finite volume effect will produce a different continuum limit value,
the slope of the points from the smaller and larger volumes agree,
consistent with errors of ${\cal O}(a^2)$.  

\begin{table}
\begin{center}
\caption{Fit parameters and $\chi^2_{\rm DF}$ for joint and separate
  fits of the FLIC, NP-improved clover and Wilson hadron masses We fit
  to an ans\"atze of the form $m_H / \sqrt{\sigma} = H_0 + H_1\,
  a\sqrt{\sigma} + H_2\, a^2 \sigma$, where the hadron, $H$, can be
  the vector meson, $V$, or the nucleon, $N$.  
\label{fits}}
\begin{ruledtabular}
  \begin{tabular}{cccccc}
        & FLIC & NP Clover &
        \multicolumn{2}{c}{Wilson} & \\
        $V_0$ & $V_2$     & $V_2$    & $V_1$     & $V_2$ & $\chi^2_{\rm DF}$
    \\ \hline 
    2.324(24) & 0.18(23)  & 0.69(18) & -1.78(20) & 0.17(43)  & 0.96  \\
    2.317(18) & 0.24(18)  & 0.74(14) & -1.71(7)  & 0         & 0.87  \\
    2.320(25) & 0.22(24)  & 0.72(19) &           &           & 0.76  \\
    2.291(34) & 0.47(32)  &          &           &           & 0.40  \\
 \hline
        $N_0$ & $N_2$     & $N_2$    & $N_1$     & $N_2$ & $\chi^2_{\rm DF}$   
    \\ \hline
    3.415(55) & -0.05(56) & 0.81(41) & -2.18(45) & 0.34(97) & 1.88  \\
    3.402(39) &  0.07(42) & 0.90(31) & -2.04(15) & 0        & 1.69  \\
    3.402(50) &  0.08(51) & 0.90(38) &           &          & 1.52  \\
    3.335(53) &  0.71(53) &          &           &          & 0.39  \\
\end{tabular}
\end{ruledtabular}
\end{center}
\vspace{-0.5cm}
\end{table}

Focusing on simulation results from physical volumes with extents
$\sim 2$ fm and larger, we perform a simultaneous fit of the FLIC,
NP-improved clover and Wilson fermion action results.  The fits are
constrained to have a common continuum limit and assume errors are
${\cal O} (a^2)$ for FLIC and NP-improved clover actions and ${\cal O}
(a)$ for the Wilson action.  
To obtain a data point at our fine lattice spacing, $a^2 \sigma \sim
0.045$, we use the observed finite volume effect at $a^2 \sigma \sim
0.075$ to correct the point at $a^2 \sigma \sim 0.045$ as illustrated
by the open triangles.
The results from these fits are given in Table~\ref{fits}.  In light
of the different gauge actions used in the analyses and the fact that
the lattice volumes considered are not perfectly matched throughout
all the simulations, an acceptable $\chi^2$ per degree of freedom is
obtained for both the nucleon and $\rho$-meson fits.

To assess the sensitivity of our results on the number of smearing
sweeps, we perform a second calculation with only four smearing sweeps
for our lattice having the coarsest lattice spacing.  This result is
indicated by the solid diamond symbol offset to the right for clarity.
These two results at $a^2 \sigma \sim 0.135$ reveal an insensitivity
to the number of APE-smearing sweeps used in constructing the
irrelevant operators of the FLIC fermion action.  This insensitivity
suggests that one could define the FLIC action in terms of a fixed
number of APE smearing sweeps independent of the lattice spacing.
Upon taking the continuum limit the smearing radius would still tend
to zero as required.

In conclusion, the use of fat links in the irrelevant operators of the
FLIC fermion action provides a new form of nonperturbative ${\cal
O}(a)$ improvement without the need of nonperturbative fine tuning.
In addition, the ${\cal O} (a^2)$ errors are small for this action.
FLIC fermions display nearly perfect scaling, providing near continuum
limit results at finite lattice spacing.

\vspace*{0.3cm} We thank the Australian Partnership for Advanced
Computing (APAC) and the Australian National Computing Facility for
Lattice Gauge Theory for generous grants of supercomputer time which
have enabled this project.  This work is supported by the Australian
Research Council.


\end{document}